\documentclass[aps,prl,twocolumn,superscriptaddress,showkeys]{revtex4-2}

\usepackage{graphicx}

\usepackage{bm}
\usepackage{xcolor}
\usepackage[colorlinks=true,citecolor=blue,linkcolor=blue,urlcolor=blue]{hyperref}
\usepackage{url}
\usepackage{bookmark}
\usepackage{amsmath}
\usepackage{xspace}
\usepackage{pdfpages}

\makeatletter
\AtBeginDocument{\let\LS@rot\@undefined}
\makeatother

\newcommand{\be}{\begin{equation}}
\newcommand{\ee}{\end{equation}}

\def\bea{\begin{eqnarray}}
\def\eea{\end{eqnarray}}

\def\vq{{\bf q}}

\def\vk{{\bf k}}

\def\tjv{$t$-$J$-$V$\xspace}

\def\ii{\mathrm{i}}
\def\lN{large-{\it N}\xspace}

\begin{document}

\title{Strongly correlated model of acousticlike plasmons persisting across the phase diagram of cuprate superconductors}

\author{Luciano Zinni}
\affiliation{Facultad de Ciencias Exactas, Ingenier\'{\i}a y Agrimensura (UNR-CONICET), Avenida Pellegrini 250, 2000 Rosario, Argentina}
\author{Hiroyuki Yamase}
\email{yamase.hiroyuki@nims.go.jp}
\affiliation{Research Center of Materials Nanoarchitectonics (MANA), National Institute for Materials Science (NIMS), Tsukuba 305-0047, Japan}
\author{Matthias Hepting}
\email{m.hepting@fkf.mpg.de}
\affiliation{Max-Planck-Institute for Solid State Research, Heisenbergstraße 1, 70569 Stuttgart, Germany}
\author{Mat\'{\i}as Bejas}
\affiliation{Facultad de Ciencias Exactas, Ingenier\'{\i}a y Agrimensura and Instituto de F\'{\i}sica Rosario (UNR-CONICET), Avenida Pellegrini 250, 2000 Rosario, Argentina}
\author{Andr\'es Greco}
\email{agreco@fceia.unr.edu.ar}
\affiliation{Facultad de Ciencias Exactas, Ingenier\'{\i}a y Agrimensura and Instituto de F\'{\i}sica Rosario (UNR-CONICET), Avenida Pellegrini 250, 2000 Rosario, Argentina}

\date{\today}

\begin{abstract}
Layered two-dimensional electron systems exhibit both optical and acousticlike plasmons around the Brillouin-zone center. In the layered cuprate La$_{2-x}$Sr$_x$CuO$_4$, resonant inelastic x-ray scattering (RIXS) has detected corresponding acousticlike plasmons in a low-energy regime comparable to that of other collective excitations associated with distinct regions of the cuprate phase diagram. This overlap in energy scale raises the question of whether the acousticlike plasmons are significantly influenced by phase-specific electronic phenomena, including the pseudogap, charge and spin order, superconductivity, and strange-metal behavior. Here we show that a single parameter set of the layered $t$-$J$-$V$ model, which incorporates strong correlations and the long-range Coulomb interaction $V$, consistently describes the acousticlike plasmon dispersion across all currently available RIXS data from the underdoped to the heavily overdoped regime. This transferability of a single parameter set exceeds that of earlier theoretical descriptions and supports a picture in which strong correlations persist into the heavily overdoped regime, while the collective plasmon mode exhibits only limited sensitivity to the phase-specific electronic phenomena that distinguish different regions of the phase diagram.
\end{abstract}

\maketitle

\textit{Introduction.} 
High-temperature cuprate superconductors exhibit a rich phase diagram as a function of hole doping and temperature, including antiferromagnetism, the pseudogap, spin and charge order, $d$-wave superconductivity, the strange-metal regime, and Fermi-liquid-like transport behavior~\cite{keimer15}. Characterizing how the associated low-energy excitations evolve across this phase diagram has been a major objective of both theory and experiment, since these excitations encode information about the underlying electronic states and the extent to which different ordering tendencies compete or coexist~\cite{lee06,scalapino12,dai01,devereaux07,damascelli03,fink01,fischer07,agterberg20,basov05}.

A powerful experimental technique to probe the low-energy spin and charge dynamics in cuprates across a wide energy and momentum range is resonant inelastic x-ray scattering (RIXS)~\cite{ament11,degroot24,mitrano24}. Specifically, RIXS at the Cu $L_3$-edge can map out dispersive spin excitations, commonly referred to as paramagnons in doped metallic cuprates~\cite{braicovich10,letacon11,dean13a,letacon13,dean13,minola15,monney16,minola17,hameed25}, which were first found by inelastic neutron scattering to emanate from the Brillouin-zone (BZ) corner $(\pi,\pi)$~\cite{birgeneau06}. Notably, the paramagnon mode retains significant spectral weight deep into the overdoped regime~\cite{wakimoto15}, although a crossover from collective spin dynamics to a regime more closely resembling individual particle-hole excitations occurs around optimal doping~\cite{monney16,minola17,hameed25}. Charge order with an incommensurate wavevector has been observed by RIXS in parts of the underdoped phase diagram, whereas charge-density fluctuations may extend over a broader doping range~\cite{ghiringhelli12,chang12,achkar12}. RIXS results further indicate that these dynamic charge correlations disappear progressively upon overdoping~\cite{tam26}. A recent resonant x-ray diffraction study also proposed the emergence of distinct charge order with a different wavevector and out-of-plane correlation in the heavily overdoped regime~\cite{li23}, although other experiments suggest that a similar diffraction signal could instead arise from  
oxygen vacancy ordering~\cite{hameed25a}. 
The doping evolution of both the paramagnon excitation and charge order has also been extensively investigated theoretically~\cite{zeyher13,benjamin14,jia14,zafur24,bejas12,allais14,meier14,wang14,atkinson15,yamakawa15,mishra15,zeyher18}.

A distinct type of charge excitation has been observed by Cu $L_3$-, O $K$-, and Cu $K$-edge RIXS in the vicinity of the BZ center~\cite{ishii05,wakimoto13,ishii17,hepting18,lin20,nag20,singh22a,hepting22,hepting23,nakata25,bejas24,nag24}. Although the origin was initially controversial~\cite{ishii14,wslee14}, it is now established that these excitations correspond to acousticlike plasmons, characteristic of layered cuprates~\cite{hepting18,lin20,nag20,singh22a,hepting22,hepting23,nakata25,bejas24,nag24}. This type of plasmons emerges in the presence of long-range Coulomb interaction in a layered structure, while their dispersion is further shaped by interlayer electron hopping. Specifically, the generic plasmon spectrum in a layered electron system depends not only on the in-plane momentum transfer $\vq_{\parallel}$, but also on the out-of-plane momentum $q_z$~\cite{fetter73,grecu73,apostol75,grecu75}. For $q_z=0$, one obtains the optical plasmon branch with a conventional quadratic dispersion, observed in early electron-energy loss spectroscopy studies~\cite{nuecker89,romberg90}. In the presence of interlayer hopping $t_z$~\cite{greco16,hepting22}, this minimum is shifted to finite energy, rendering the branches acousticlike rather than purely acoustic. For the theoretical description of plasmons in cuprates, a variety of approaches has been employed \cite{greco16,ruvalds87,bill03,markiewicz07a,markiewicz08,zaanen19,fidrysiak21,boyd22,gabriele22,silkin23,sellati25,yamase25}. 

\begin{figure}[tb]
   \centering
   \includegraphics[width=\linewidth]{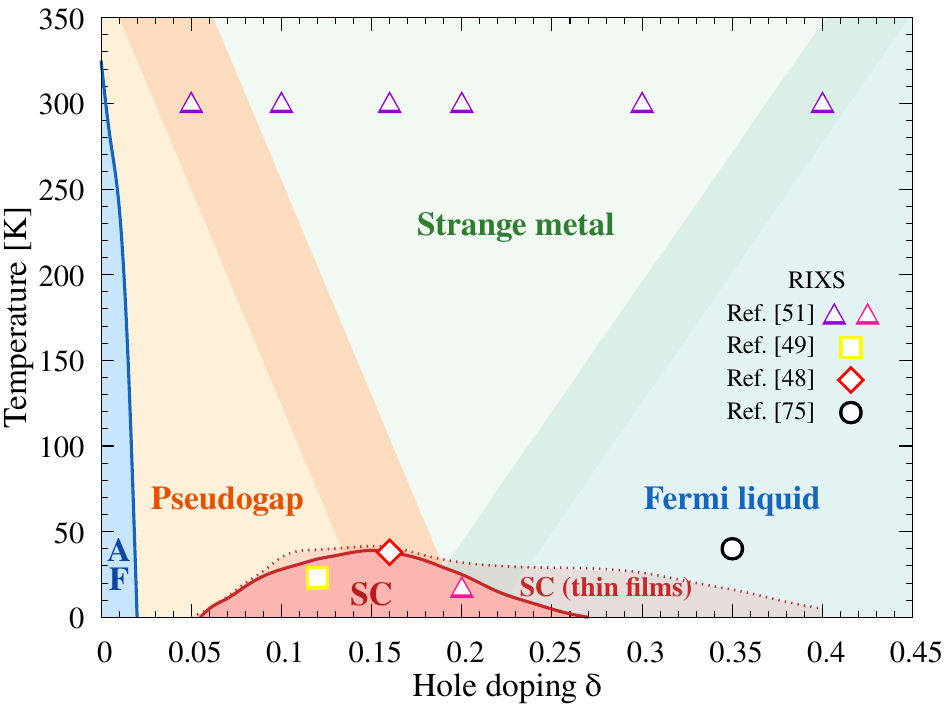} 
   \caption{Schematic phase diagram of La$_{2-x}$Sr$_x$CuO$_4$ as a function of hole doping and temperature, compiled from Refs.~\cite{takagi89,keimer92,niedermayer98,hashimoto07,cooper09,hepting23}. Colored regions indicate the antiferromagnetic (AF) phase, the pseudogap regime (orange) with its crossover region (dark orange), the strange-metal regime (green) with its crossover region (dark green), and the regime of Fermi-liquid-like transport behavior (blue). The superconducting (SC) dome is shown by the solid red line for bulk samples and by the dashed red line for thin films. Symbols mark the locations in the phase diagram of RIXS measurements: purple triangles correspond to O $K$-edge RIXS on thin films in Ref.~\cite{hepting23}, the yellow square to O $K$-edge RIXS on a single crystal in Ref.~\cite{singh22a}, the red diamond to O $K$-edge RIXS on a single crystal in Ref.~\cite{nag20}, the pink triangle to Cu $K$-edge RIXS on a single crystal in Ref.~\cite{hepting23}, and the black circle to O $K$-edge RIXS on a thin film in Ref.~\cite{dean26}. 
 }
   \label{fig:phase-diagram}
\end{figure}

RIXS measurements on the prototypical cuprate superconductor La$_{2-x}$Sr$_x$CuO$_4$ (LSCO)~\cite{nag20,singh22a,hepting23}, where the Sr substitution $x$ is equivalent to the hole doping $\delta$, have reported acousticlike plasmon excitations at various locations in the phase diagram (Fig.~\ref{fig:phase-diagram}). These measurements span the pseudogap, superconducting, strange-metal, and Fermi-liquid regimes, and were performed on both thin-film and single-crystal samples at the O $K$- and Cu $K$-edges. Most recently, a detailed high-resolution RIXS data set was reported for heavily overdoped LSCO with $\delta=0.35$~\cite{dean26}, measured at $T=40$~K deep in the Fermi-liquid regime, where spin fluctuations are strongly damped and the pseudogap as well as superconductivity are absent~\cite{proust02,plate05,vignolle08,letacon13,dean13,minola15,wakimoto15}.

In this Letter, we compile the available RIXS data on LSCO between $\delta = 0.05$ and $0.40$~\cite{nag20,singh22a,hepting23,dean26} and model the experimentally reported plasmon dispersion using the layered $t$-$J$-$V$ model, which extends the standard two-dimensional $t$-$J$ model to incorporate both interlayer hopping and long-range Coulomb interaction $V$ in a form that respects the lattice structure~\cite{greco16}. Notably, we find that a single microscopic parameter set of the $t$-$J$-$V$ model, previously determined for optimally doped LSCO ($\delta=0.16$)~\cite{hepting22}, provides a consistent description of all reported plasmon dispersions across the phase diagram when only adjusting the hole doping $\delta$ and the temperature $T$. This degree of transferability and consistency surpasses that of other theoretical descriptions of the plasmon dispersion in cuprates, including the random phase approximation (RPA) models~\cite{markiewicz07a,markiewicz08,dean26}. It further suggests that strong correlations, which are inherently included in the $t$-$J$-$V$ model, persist in the overdoped regime and that, unlike many other physical observables~\cite{timusk99}, the acousticlike plasmon remains a robust collective mode across different hole doping and temperature regions, with its dispersion showing only limited sensitivity to the electronic states in the cuprate phase diagram.

\begin{figure*}[tb]
   \centering
   \includegraphics[width=\linewidth]{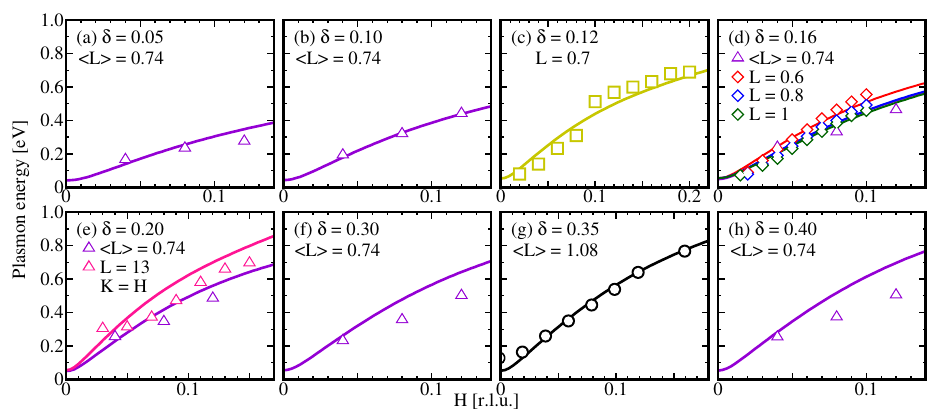} 
   \caption{In-plane plasmon dispersion. (a-h) Open symbols are plasmon energies determined in RIXS measurements on LSCO with hole doping $\delta = 0.05, 0.10, 0.12, 0.16, 0.20, 0.30, 0.35$, and $0.40$, respectively. Measurements were performed by varying the momentum transfer along the in-plane $H$ direction, while the out-of-plane momentum $L$ was fixed. For experiments where the out-of-plane momentum was not strictly fixed during $H$ variation, the label $\langle L\rangle$ denotes the average out-of-plane momentum. The data labeled with $H=K$ in panel (e) were measured for in-plane momentum transfer along the diagonal direction, whereas in all other measurements $K=0$ was fixed. Solid lines are plasmon dispersions calculated with the $t$-$J$-$V$ model, using the same set of microscopic parameters (see text) determined for LSCO with $\delta = 0.16$ in Ref.~\cite{hepting22}, while adjusting only $\delta$ and the temperature according to the corresponding experiment. RIXS data in panel (a) and (b) are from Ref.~\cite{hepting23}, in (c) from Ref.~\cite{singh22a}, in (d) from Refs.~\cite{hepting23,nag20}, in (e) from Ref.~\cite{hepting23}, in (f) from Ref.~\cite{hepting23}, in (g) from Ref.~\cite{dean26}, and in (h) from Ref.~\cite{hepting23}.        
}
   \label{fig:Hscan}
\end{figure*}

\textit{Model and formalism.} To describe plasmon excitations in LSCO, we employ the layered \tjv model on a square lattice,
\begin{align}
H =& -\sum_{i,j,\sigma} t_{ij} \tilde{c}^\dagger_{i\sigma}\tilde{c}_{j\sigma} + \sum_{\langle i,j\rangle} J_{ij}\left(\vec{S}_i \cdot \vec{S}_j - \frac{1}{4}n_i n_j\right) \nonumber \\
  &+ \frac{1}{2}\sum_{\substack{i,j \\ i \neq j}} V_{ij} n_i n_j\;,
\label{eq:hamiltonian}
\end{align}
where $\tilde{c}^\dagger_{i\sigma}$ ($\tilde{c}_{i\sigma}$) creates (annihilates) an electron with spin $\sigma$ in the Hilbert space without double occupancy. The hopping amplitudes $t_{ij}$ include nearest-neighbor ($t$) and next-nearest-neighbor ($t'$) processes within each CuO$_{2}$ plane, as well as interlayer hopping $t_z$. The exchange interaction $J_{ij}=J$ acts between nearest neighbors within the planes, and $V_{ij}$ denotes the long-range Coulomb interaction. $n_i=\sum_\sigma \tilde{c}^\dagger_{i\sigma} \tilde{c}_{i\sigma}$ and $\vec{S}_i$ are the electron density and spin operators, respectively.
The non-double-occupancy constraint is treated via a path-integral formulation of Hubbard operators within a large-$N$ expansion (technical details are given in the Supplemental Material~\cite{SM}; see also Refs.~\cite{hubbard63,faddeev88,bejas17,prelovsek99} therein). The quasiparticle dispersion $\varepsilon_\vk = \varepsilon^\parallel_\vk + \varepsilon^\perp_\vk$ is given by
\begin{align}
\varepsilon^\parallel_\vk =& -2\left(t \frac{\delta}{2} + \Delta\right)(\cos k_x + \cos k_y) \nonumber \\ 
                           &- 4t'\frac{\delta}{2}\cos k_x \cos k_y - \mu\;,
\label{eq:disp_parallel}
\end{align}
\be
\varepsilon^\perp_\vk = -2t_z\frac{\delta}{2}(\cos k_x - \cos k_y)^2 \cos k_z\,, 
\label{eq:disp_perp}
\ee
where $\delta/2$ originates from strong correlations and controls the effective bandwidth and interlayer hopping, and the bond field $\Delta$ and chemical potential $\mu$ are determined self-consistently~\cite{SM}. The long-range Coulomb interaction consistent with a layered square lattice~\cite{becca96} takes the form
\be
V(\vq) = \frac{V_c}{A(q_x, q_y) - \cos q_z}\;,
\label{eq:coulomb}
\ee
with $V_c = e^2d/(2\varepsilon_\perp a^2)$ and $A(q_x, q_y) = \alpha(2 - \cos q_x - \cos q_y) + 1$. Here $\alpha = \tilde{\varepsilon}/(a/d)^2$ with $\tilde{\varepsilon} = \varepsilon_\parallel/\varepsilon_\perp$, where $\varepsilon_\parallel$ and $\varepsilon_\perp$ are the dielectric constants parallel and perpendicular to the CuO$_2$ planes, respectively. The lattice constant within the planes is denoted by $a$, the interlayer spacing by $d$, and $e$ is the electron charge. Within the \tjv framework, the charge-charge correlation function is obtained at leading order as
\be
\chi_c(\vq, \ii\omega_n) = N\left(\frac{\delta}{2}\right)^2 D_{11}(\vq, \ii\omega_n)\;,
\label{eq:chi}
\ee
where $D_{11}$ is the $(1,1)$ element of the $6\times 6$ dressed bosonic propagator obtained from the Dyson equation~\cite{SM}, and $\vq$ and $\ii\omega_n$ are the momentum transfer and bosonic Matsubara frequency, respectively. After analytical continuation $\ii\omega_n \to \omega + \ii\Gamma$ and setting the physical value $N=2$, we compute ${\rm Im}\chi_c(\vq,\omega)$, which is directly compared with RIXS intensity maps.

\textit{Results.} 
We now examine whether a coherent description of the plasmon dispersion can be achieved across the entire doping range measured with RIXS. To this end, we adopt the parameter set previously introduced for optimally doped LSCO ($\delta=0.16$) in Ref.~\cite{hepting22}: $t'/t=-0.2$, $t_z/t=0.01$, $V_c/t=31$, $\alpha=3.5$, $\Gamma/t=0.1$, and $J/t=0.3$, with $t/2=0.35$ eV. Since in the \lN formalism the original Hamiltonian hopping $t$ is scaled to $t/N$, and $N=2$ is set at the end of the calculation~\cite{bejas12,greco16}, the natural unit of energy is $t/2$, consistent with the hopping expected for cuprates (see Supplemental Material~\cite{SM}).

Figure~\ref{fig:Hscan} compares the measured in-plane plasmon dispersions for eight hole dopings between $\delta = 0.05$ and $0.40$ (open symbols) with the corresponding \tjv calculations (solid lines). For each calculated dispersion, only the hole doping, the temperature, and the momentum transfer $(H,K,L)$ are adjusted to match the experimental conditions. The momentum coordinates are expressed in reciprocal lattice units (r.l.u.).

Overall, the calculated in-plane dispersions are in good agreement with the RIXS data throughout the full doping range. This includes cases in which the dispersion at fixed $\delta$ was measured for different values of $L$ [Figs.~\ref{fig:Hscan}(d),(e)], as well as measurements performed at the O $K$- and the Cu $K$-edges, and on both thin-film and single-crystal samples (see also Fig.~\ref{fig:phase-diagram}). In the overdoped regime, however, the RIXS data for $\delta = 0.30$ and $0.40$ reported in Ref.~\cite{hepting23} deviate somewhat from the calculated \tjv dispersions, particularly at large $H$. By contrast, the high-resolution RIXS data for $\delta = 0.35$ reported in Ref.~\cite{dean26} agree very well with the calculated dispersion at every finite $H$. This point will be further discussed below.

\begin{figure}[tb]
   \centering
   \includegraphics[width=\linewidth]{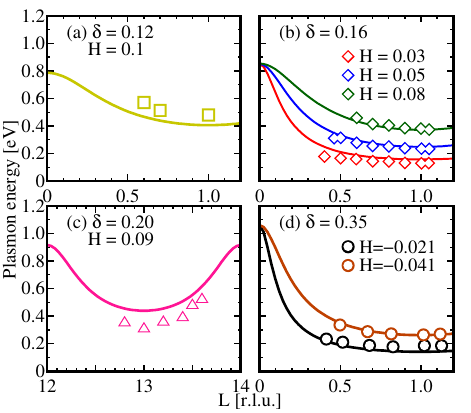} 
    \caption{Out-of-plane plasmon dispersion. (a-d) Open symbols are plasmon energies determined in RIXS measurements on LSCO with hole dopings $\delta = 0.12, 0.16, 0.20$, and 0.35, respectively. Measurements were performed by varying the momentum transfer along the out-of-plane $L$ direction, while the in-plane momentum $H$ was fixed. Solid lines are $t$-$J$-$V$ model calculations, using the same set of microscopic parameters determined for LSCO with $\delta = 0.16$ in Ref.~\cite{hepting22}, while adjusting only $\delta$ and the temperature. RIXS data in panel (a) are from Ref.~\cite{singh22a}, in (b) from Ref.~\cite{nag20}, in (c) from Ref.~\cite{hepting23}, and in (d) from Ref.~\cite{dean26}. 
}
   \label{fig:Lscan}
\end{figure}

Figure~\ref{fig:Lscan} provides the complementary comparison for the out-of-plane plasmon dispersion. Similarly to the in-plane dispersion, the overall agreement between experiment and theory is satisfactory for all measured hole dopings, $\delta = 0.12$, $0.16$, $0.20$, and $0.35$, with small deviations occurring for $\delta = 0.12$ and $0.20$ [Figs.~\ref{fig:Lscan}(a),(c)]. 

Taken together, Figs.~\ref{fig:Hscan} and \ref{fig:Lscan} illustrate that the same microscopic parameter set captures the main features of the three-dimensional plasmon dispersion reported in various RIXS experiments, without tuning model parameters individually for each doping or introducing phase-specific corrections. This result is especially significant in the overdoped regime, where the continued relevance of strong correlations has been debated~\cite{nag24,hepting23,nakamae03,letacon13,dean13,minola15,wakimoto15}, signaling that a strongly correlated framework retains predictive power even in that limit.

\textit{Discussion.} 
Notably, our results differ qualitatively from those obtained within a random-phase approximation (RPA) treatment. While RPA can reproduce the plasmon dispersion at a given doping level after introducing ad hoc renormalized band parameters~\cite{nag24}, the same parameter set does not remain valid when the hole doping is changed. Specifically, as the hole doping moves away from the value for which the RPA calculation was adjusted, the discrepancy with experiment becomes increasingly pronounced (see Supplemental Material~\cite{SM}). By contrast, the layered \tjv model captures the plasmon dispersion across the available doping range using a single microscopic parameter set.

This difference reflects how doping enters the two descriptions. In the \tjv framework, the quasiparticle dispersion contains the correlation-induced renormalization factor $\delta/2$ in Eqs.~\eqref{eq:disp_parallel} and \eqref{eq:disp_perp}, which controls both the effective bandwidth and interlayer hopping. The plasmon dispersion, however, is not set by the band structure alone. It emerges from the poles of the dressed charge propagator $D_{11}(\vq,\omega)$, whose doping dependence is governed by the Dyson equation and the bosonic self-energy $\Pi_{ab}(\vq,\omega)$, which carries a nontrivial doping dependence beyond the $\delta/2$ factor in the quasiparticle dispersion (for more details, see Supplemental Material~\cite{SM}). The fact that the same \tjv parameter set describes the plasmon dispersion across the full doping range thus demonstrates that the collective charge response encoded in $D_{11}(\vq,\omega)$ evolves consistently. This indicates that the \tjv framework provides a more unified description of cuprate plasmons than an RPA treatment, even if RPA can be adjusted to reproduce individual cases.

The ability of a single \tjv parameter set to account for the plasmon dispersion across the cuprate phase diagram has important physical implications. It reveals that the acousticlike plasmon is not primarily controlled by the degrees of freedom that distinguish the pseudogap, strange-metal, and Fermi-liquid regimes, even though the acousticlike branches disperse down to energies comparable to those of the pseudogap, charge-density-wave fluctuations, spin excitations, and the superconducting gap. In particular, the measured plasmon dispersion remains quantitatively accounted for as the system evolves from a regime with strong antiferromagnetic correlations, pseudogap, and superconducting ground state to one in which these features are absent or substantially weakened. This indicates that, for the description of the currently available RIXS data, the leading-order \tjv formalism 
remains adequate without explicit inclusion of pseudogap formation, superconducting pairing, or charge ordering.

Nonetheless, as mentioned above, the RIXS data for $\delta = 0.30$ and $0.40$ reported in Ref.~\cite{hepting23} deviate somewhat from the calculated \tjv dispersion, especially at large $H$ [Figs.~\ref{fig:Hscan}(f),(h)], whereas the more recent $\delta = 0.35$ data of Ref.~\cite{dean26} are captured very well [Fig.~\ref{fig:Hscan}(g)]. In Ref.~\cite{hepting23}, this discrepancy was discussed in terms of three possible scenarios: (i) enhanced structural disorder in overdoped samples, (ii) an overestimation of correlation effects within the \tjv model at high doping where the non-double-occupancy constraint and nearest-neighbor exchange become less restrictive, and (iii) the increasing relevance of additional orbital degrees of freedom, including Cu $3d_{3z^2-r^2}$ and $4s$. The excellent agreement with the more comprehensive and higher-resolution data of Ref.~\cite{dean26}, however, suggests that the latter two scenarios do not need to be invoked given the newly available $\delta = 0.35$ data, and that the \tjv framework is capable of describing the plasmon dispersion even in the heavily overdoped regime.

While the overall agreement shows that the \tjv framework is transferable across the available doping range, exact quantitative agreement across all compiled RIXS data should not be expected, because real samples differ in additional microscopic details beyond $\delta$. In particular, the in-plane and out-of-plane lattice constants vary substantially between $\delta = 0.05$ and $0.40$ and differ between bulk and thin film samples, which in principle should affect the Coulomb interaction in the \tjv framework through $\alpha=\tilde{\varepsilon}/(a/d)^2$. Residual deviations such as those in Figs.~\ref{fig:Hscan}(f),(h), and possibly also the mismatch in the out-of-plane scan of the single crystal in Fig.~\ref{fig:Lscan}(c), are therefore naturally attributable to sample-dependent changes and/or sample-quality issues not captured when $\delta$ is taken as the only doping-dependent input. Accordingly, they should not be interpreted as an indication for a breakdown of the \tjv description.

Furthermore, the demonstrated robustness of the \tjv description does not exclude weak modifications of the plasmon across particular electronic transitions. Specifically, optical measurements on bismuth-based cuprates have detected subtle renormalization effects on the optical plasmon across $T_c$~\cite{levallois16}. It can therefore be expected that the opening of the superconducting gap similarly affects the low-energy acousticlike plasmon branches in LSCO, although this has not yet been resolved by RIXS and may require improved energy resolution and finer temperature sampling compared to previous experiments. In addition, in layered nickelates a hardening of the plasmon energy, which might be related to stripe fluctuations, was observed~\cite{dean26}. This suggests that charge and spin stripes in the underdoped regime of LSCO could likewise weakly affect the plasmon dispersion.

\textit{Conclusion.} 
In summary, we have shown that the layered $t$-$J$-$V$ model provides a consistent account of the measured acousticlike plasmon dispersion in LSCO from the underdoped to the heavily overdoped regime using a single microscopic parameter set while varying only the hole doping $\delta$. This agreement includes the latest available RIXS data on the three-dimensional plasmon dispersion in LSCO \cite{dean26}, and therefore supports a unified description of the plasmon throughout the cuprate phase diagram. Taken together, our results indicate that the low-energy acousticlike plasmon is a robust collective charge excitation of the correlated electron system, only weakly affected by the electronic phenomena that distinguish the different regions of the cuprate phase diagram.

\textit{Acknowledgments.}
We thank M. P. M. Dean and his collaborators for insightful discussions, and for making the LSCO RIXS data from Ref.~\cite{dean26} available prior to publication. We are also grateful to M. P. M. Dean for a careful reading of an early version of the manuscript. We further thank M. Minola and C. Falter for valuable discussions. H.Y. was supported by World Premier International  Research Center Initiative (WPI), MEXT, Japan. A part of the results presented in this work was obtained by using the facilities of the CCT-Rosario Computational Center, member of the High Performance Computing National System (SNCAD, MincyT- Argentina).

\textit{Data availability.} The theoretical data generated in this study are available from the corresponding authors upon reasonable request.

\bibliography{main}


\clearpage


\section*{Supplementary material}

# Supplemental Material for "Strongly correlated model of acousticlike plasmons persisting across the phase diagram of cuprate superconductors"

Luciano Zinni,[1] Hiroyuki Yamase,[2, *] Matthias Hepting,[3, †] Matías Bejas,[4] and Andrés Greco[4, ‡]

[1]*Facultad de Ciencias Exactas, Ingeniería y Agrimensura (UNR-CONICET),*
*Avenida Pellegrini 250, 2000 Rosario, Argentina*
[2]*Research Center of Materials Nanoarchitectonics (MANA),*
*National Institute for Materials Science (NIMS), Tsukuba 305-0047, Japan*
[3]*Max-Planck-Institute for Solid State Research, Heisenbergstraße 1, 70569 Stuttgart, Germany*
[4]*Facultad de Ciencias Exactas, Ingeniería y Agrimensura and Instituto de Física Rosario (UNR-CONICET),*
*Avenida Pellegrini 250, 2000 Rosario, Argentina*

## LARGE-$N$ PATH-INTEGRAL TREATMENT OF THE $t$-$J$-$V$ MODEL

This section provides the details of the large-$N$ path-integral treatment of the $t$-$J$-$V$ Hamiltonian [Eq. (1) in the main text]. For a more comprehensive derivation of the formalism see Ref. [63]; here we summarize the key expressions needed to obtain the charge-charge correlation function $\chi_c(\mathbf{q}, \omega)$ presented in Eq. (5) in the main text.

### Effective bosonic theory

The starting point is the representation of the constrained electronic operators in terms of Hubbard $\hat{X}$-operators, which satisfy non-standard commutation rules and are defined as $\hat{X}_i^{\alpha\beta} = |i\alpha\rangle\langle i\beta|$ with $|\alpha\rangle = |0\rangle, |\uparrow\rangle, |\downarrow\rangle$ [86]. In this basis, for instance, it follows that $\tilde{c}_{i\sigma}^\dagger = X_i^{\sigma 0}$ and $\tilde{c}_{i\sigma} = X_i^{0\sigma}$. Besides, the non-double-occupancy constraint is encoded in the completeness relation $X_i^{00} + \sum_\sigma X_i^{\sigma\sigma} = 1$, and the Hamiltonian [Eq. (1) in the main text] becomes quadratic in the $\hat{X}$-operators.

A path-integral treatment based on the Faddeev-Jackiw method [87] maps this problem onto an effective field theory. After extending the spin index from 2 to $N$ and performing an expansion in powers of $1/N$, the resulting effective Lagrangian describes a theory of fermions, bosons, and their interactions. The fermionic sector yields the quasiparticle dispersion $\varepsilon_\mathbf{k}$ given in Eqs. (2) and (3) of the main text, where the hopping integrals acquire the correlation-induced renormalization factor $\delta/2$ and the bond field $\Delta$. The bosonic sector is captured by a six-component field

$$\delta X_i = (\delta R_i, \ \delta\lambda_i, \ r_i^x, \ r_i^y, \ A_i^x, \ A_i^y), \quad (A1)$$

whose components have the following physical origin. The field $\delta R_i$ parametrizes fluctuations of the on-site hole doping $\delta$ around its mean-field value through $X_i^{00} = N\frac{\delta}{2}(1 + \delta R_i)$. The field $\delta\lambda_i$ represents fluctuations of the Lagrange multiplier that enforces the non-double-occupancy constraint beyond mean field. The remaining four components arise from a Hubbard-Stratonovich decoupling of the exchange interaction $J$: the bond variable along the direction $\eta = x, y$ is written as $\Delta_i^\eta = \Delta(1 + r_i^\eta + \mathrm{i}A_i^\eta)$, so that $r_i^\eta$ and $A_i^\eta$ represent fluctuations of its real and imaginary parts, respectively. Here $\Delta$ is the static mean-field bond-order parameter, determined self-consistently together with the chemical potential $\mu$ through

$$\Delta = \frac{J}{4N_s}\sum_\mathbf{k}(\cos k_x + \cos k_y)n_F(\varepsilon_\mathbf{k}), \quad (A2)$$

$$(1 - \delta) = \frac{2}{N_s}\sum_\mathbf{k} n_F(\varepsilon_\mathbf{k}), \quad (A3)$$

where $N_s$ is the total number of lattice sites and $n_F$ is the Fermi-Dirac distribution function.

The quadratic part of the Lagrangian in $\delta X_i$ defines a bare bosonic propagator $D_{ab}^{(0)}(\mathbf{q}, \mathrm{i}\omega_n)$, whose inverse reads

$$[D_{ab}^{(0)}(\mathbf{q}, \mathrm{i}\omega_n)]^{-1} = N \begin{pmatrix} \frac{\delta^2}{2}[V(\mathbf{q}) - J(\mathbf{q})] & \frac{\delta}{2} & 0 & 0 & 0 & 0 \\ \frac{\delta}{2} & 0 & 0 & 0 & 0 & 0 \\ 0 & 0 & \frac{4\Delta^2}{J} & 0 & 0 & 0 \\ 0 & 0 & 0 & \frac{4\Delta^2}{J} & 0 & 0 \\ 0 & 0 & 0 & 0 & \frac{4\Delta^2}{J} & 0 \\ 0 & 0 & 0 & 0 & 0 & \frac{4\Delta^2}{J} \end{pmatrix}, \quad (A4)$$

with matrix indices $a, b = 1, \ldots, 6$ corresponding to the components in Eq. (A1), and $J(\mathbf{q}) = \frac{J}{2}(\cos q_x + \cos q_y)$. The overall factor of $N$ implies that $D_{ab}^{(0)}$ is $O(1/N)$. Note that $[D_{ab}^{(0)}]^{-1}$ is frequency-independent and that the long-range Coulomb interaction $V(\mathbf{q})$ [Eq. (4) of the main text] enters exclusively in the $(1,1)$ element and is responsible for the emergence of collective plasmon modes.

### Dressed propagator and bosonic self-energy

The dressed bosonic propagator is obtained through the Dyson equation

$$D_{ab}^{-1}(\mathbf{q}, i\omega_n) = [D_{ab}^{(0)}(\mathbf{q}, i\omega_n)]^{-1} - \Pi_{ab}(\mathbf{q}, i\omega_n), \quad (A5)$$

where $\Pi_{ab}$ is the $6 \times 6$ bosonic self-energy. Its explicit form is

$$\Pi_{ab}(\mathbf{q}, i\omega_n) = -\frac{N}{N_s} \sum_\mathbf{k} h_a(\mathbf{k}, \mathbf{q}, \varepsilon_\mathbf{k} - \varepsilon_{\mathbf{k-q}})$$

$$\times \frac{n_F(\varepsilon_{\mathbf{k-q}}) - n_F(\varepsilon_\mathbf{k})}{i\omega_n - \varepsilon_\mathbf{k} + \varepsilon_{\mathbf{k-q}}} h_b(\mathbf{k}, \mathbf{q}, \varepsilon_\mathbf{k} - \varepsilon_{\mathbf{k-q}})$$

$$- \delta_{a1}\delta_{b1} \frac{N}{N_s} \sum_\mathbf{k} \frac{\tilde{\varepsilon}_\mathbf{k} - \tilde{\varepsilon}_{\mathbf{k-q}}}{2} n_F(\varepsilon_\mathbf{k}), \quad (A6)$$

where $\tilde{\varepsilon}_\mathbf{k}$ is equal to $\varepsilon_\mathbf{k}$ with $\Delta = 0$, and the six-component vertex is

$$h_a(\mathbf{k}, \mathbf{q}, \nu) = \left\{ \frac{2\varepsilon_{\mathbf{k-q}} + \nu + 2\mu}{2} \right.$$
$$+ 2\Delta \left[ \cos\left(k_x - \frac{q_x}{2}\right) \cos\frac{q_x}{2} + \cos\left(k_y - \frac{q_y}{2}\right) \cos\frac{q_y}{2} \right];$$
$$1; \ -2\Delta \cos\left(k_x - \frac{q_x}{2}\right); \ -2\Delta \cos\left(k_y - \frac{q_y}{2}\right);$$
$$\left. 2\Delta \sin\left(k_x - \frac{q_x}{2}\right); \ 2\Delta \sin\left(k_y - \frac{q_y}{2}\right) \right\}. \quad (A7)$$

A distinctive feature of this vertex is that it does not originate solely from the interactions in the Hamiltonian: the non-trivial algebra of the Hubbard $\hat{X}$-operators and the enforcement of the local constraint both contribute to its structure. In particular, the first component ($a = 1$) carries an explicit dependence on the fermionic frequency through $\nu$, while the remaining five components depend only on momenta. The out-of-plane momenta $k_z$ and $q_z$ enter exclusively through $\varepsilon_{\mathbf{k-q}}$ in the $a = 1$ component; the other components involve only the in-plane momentum $\mathbf{q}_\parallel$.

### Charge-charge correlation function

The full propagator $D_{ab}(\mathbf{q}, i\omega_n)$ encodes the complete spectrum of charge fluctuations in the $t$-$J$-$V$ model [88].

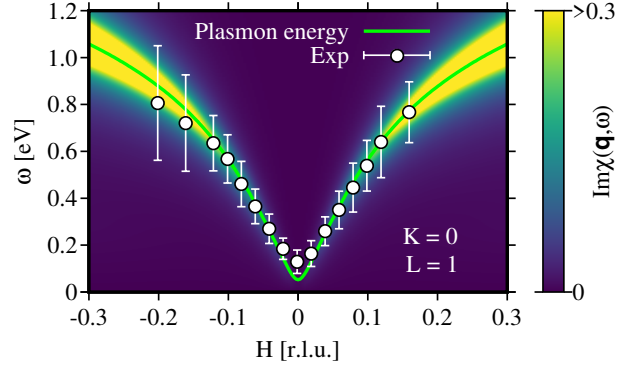

FIG. S1. Energy-momentum maps of Im $\chi_c(\mathbf{q}, \omega)$ for $\delta = 0.35$ computed within the $t$-$J$-$V$ model by employing the same parameter set previously determined for optimally doped LSCO ($\delta = 0.16$) in Ref. [50]. White dots represent the experimental plasmon energies extracted from Ref. [75] on a non-superconducting LSCO sample at $\delta = 0.35$; the bars are the plasmon width. Green lines indicate the theoretical plasmon dispersion.

Its structure reflects two physically distinct sectors: the $2 \times 2$ charge sector ($a, b = 1, 2$), which describes usual charge fluctuations, including plasmon modes; and the $4 \times 4$ bond sector ($a, b = 3$–$6$) originated from the $J$ term. To obtain plasmons it is sufficient to focus only on the $2 \times 2$ charge sector [88].

The standard charge-charge correlation function $\chi_c(\mathbf{r}_i - \mathbf{r}_j, \tau) = \langle T_\tau n_i(\tau) n_j(0) \rangle$ is written in the large-$N$ framework as

$$\chi_c(\mathbf{r}_i - \mathbf{r}_j, \tau) = \frac{1}{N} \sum_{pq} \langle T_\tau X_i^{pp}(\tau) X_j^{qq}(0) \rangle. \quad (A8)$$

Using the completeness condition $\sum_p X_i^{pp} = N/2 - X_i^{00}$ and $X_i^{00} = N\frac{\delta}{2}(1 + \delta R_i)$, the charge-charge correlation function can be expressed in Fourier space as

$$\chi_c(\mathbf{q}, i\omega_n) = N \left(\frac{\delta}{2}\right)^2 D_{11}(\mathbf{q}, i\omega_n). \quad (A9)$$

While $\chi_c$ is formally of $O(1)$, its evaluation requires all $O(1/N)$ contributions entering $D_{11}$ through the Dyson equation (A5). The retarded response is obtained via the analytical continuation $i\omega_n \to \omega + i\Gamma$ and by setting $N = 2$. The broadening parameter $\Gamma > 0$ effectively incorporates both finite experimental resolution and intrinsic incoherent scattering processes [89]. The resulting spectral function Im $\chi_c(\mathbf{q}, \omega)$ can be compared directly to RIXS intensity maps. We note that in the large-$N$ formalism the hopping parameters and interaction strengths are scaled by $1/N$ [35]. When setting $N$ to its physical value for comparison with experiments, the natural unit of energy becomes $t/2$. This is the origin of the energy scale $t/2 = 0.35$ eV employed in the main text.

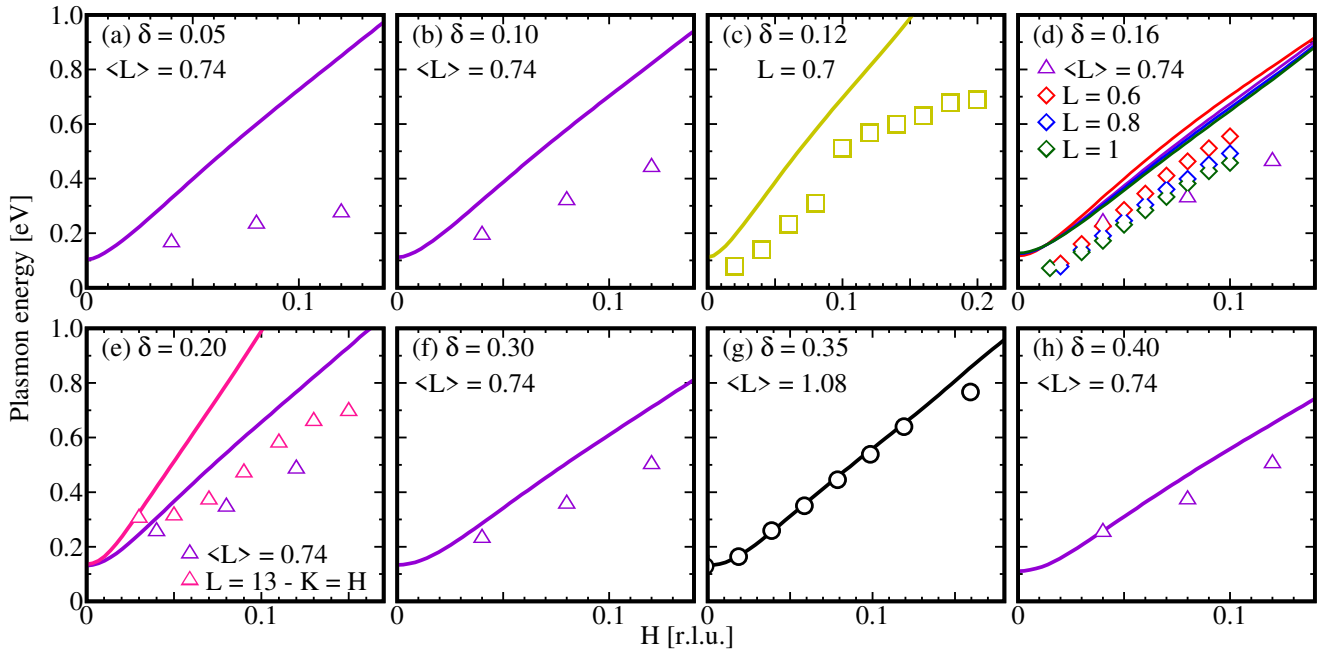

FIG. S2. In-plane plasmon dispersion in LSCO with dispersions calculated within RPA using a single parameter set fixed to the $\delta = 0.35$ data. Open symbols are the same plasmon energies determined by RIXS as shown in Fig. 2 of the main text. Solid lines are RPA dispersions obtained with the parameter set reported in Ref. [75], which was introduced to describe the $\delta = 0.35$ data and is applied here unchanged to all dopings. While this parameter set reproduces the $\delta = 0.35$ dispersion [panel (g)], it does not provide a comparably consistent description across the full doping range.

## PLASMON DISPERSION OVER THE FULL $H$-SCAN AT $\delta = 0.35$

In Fig. 2(g) of the main text, the plasmon dispersion at $\delta = 0.35$ is shown for positive values of $H$ so as to directly compare with all available RIXS data from other experiments. Figure S1 extends this comparison to the full $H$-scan reported in Ref. [75], demonstrating that the $t$-$J$-$V$ model captures well the experimental dispersion over the entire range of in-plane momenta, including negative $H$. As for all other dopings across the phase diagram, this agreement is achieved without any parameter adjustment: the same microscopic parameter set previously determined for optimally doped LSCO ($\delta = 0.16$) is employed throughout.

## RANDOM PHASE APPROXIMATION ANALYSIS

We performed a complementary analysis within the standard random phase approximation (RPA). The RPA charge susceptibility is given by

$$\chi^{\mathrm{RPA}}(\mathbf{q},\omega) = \frac{\chi_0(\mathbf{q},\omega)}{1 - V(\mathbf{q})\chi_0(\mathbf{q},\omega)}, \quad (A10)$$

where $V(\mathbf{q})$ is the long-range Coulomb interaction [Eq. (4) of the main text] and $\chi_0(\mathbf{q},\omega)$ is the Lindhard function computed from a tight-binding dispersion. In contrast to the $t$-$J$-$V$ approach, the RPA employs a bare electronic dispersion $E_{\mathbf{k}} = E_{\mathbf{k}}^{\parallel} + E_{\mathbf{k}}^{\perp}$, with

$$E_{\mathbf{k}}^{\parallel} = -2t(\cos k_x + \cos k_y) - 4t'\cos k_x \cos k_y - \mu, \quad (A11)$$

and

$$E_{\mathbf{k}}^{\perp} = -2t_z(\cos k_x - \cos k_y)^2 \cos k_z, \quad (A12)$$

where hopping amplitudes are treated as doping-independent constants [54], and the charge-fluctuation vertex reduces to unity—in contrast to the more complex vertex of the $t$-$J$-$V$ formalism, which encodes the non-canonical algebra of Hubbard operators and the non-double-occupancy constraint.

It is well established that RPA with bare band parameters yields a doping dependence of the plasmon energy opposite to experiment [51], and that a quantitative RPA description at a given doping requires the use of ad hoc renormalized band parameters [54]. Figure S2 shows the RPA dispersion obtained with the parameters of Ref. [75] ($t = 0.39$ eV, $t'/t = -0.3$, $t_z/t = 0.017$, $V_c/t = 15$, $\alpha = 3.5$, and $\Gamma/t = 0.013$), which were determined to fit the data at $\delta = 0.35$. While the RPA reproduces the plasmon dispersion at that doping, it systematically overestimates the plasmon energy at lower dopings and fails to capture the momentum dependence across the phase


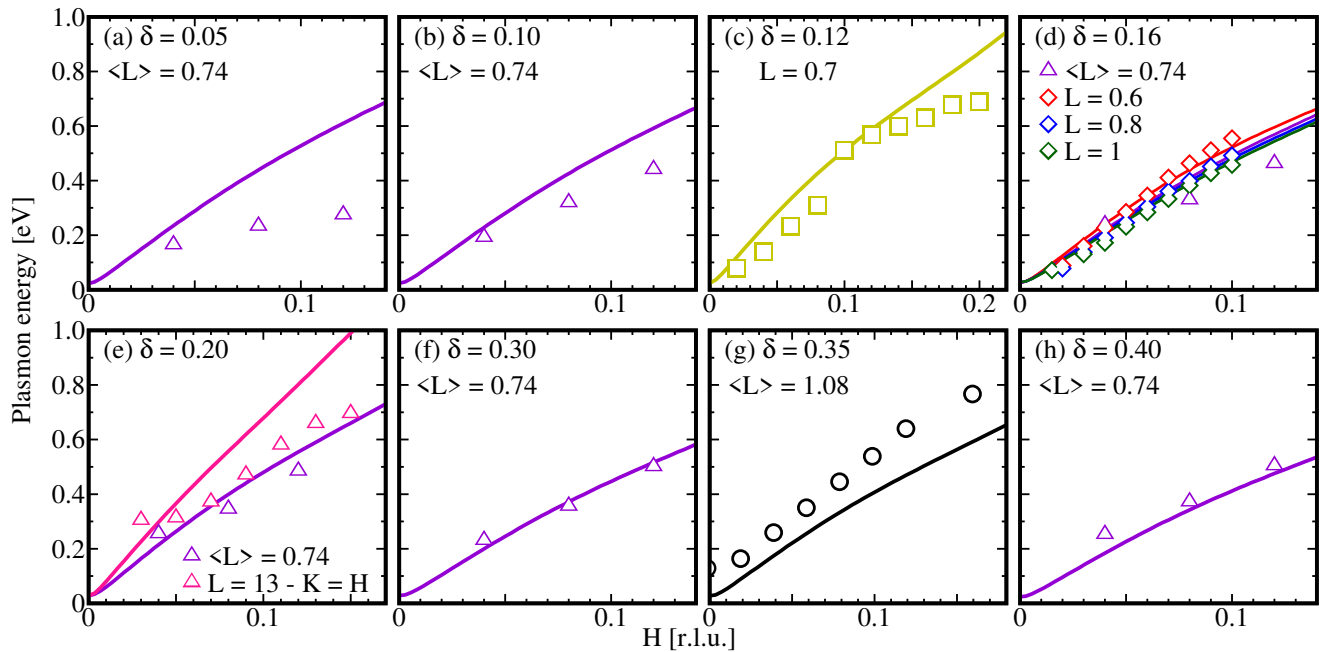

FIG. S3. In-plane plasmon dispersion in LSCO with dispersions calculated for RPA parameters adjusted for $\delta = 0.16$. (a-h) Open symbols are the same plasmon energies determined by RIXS that are given in Fig. 2 of the main text. Solid lines are plasmon dispersions calculated by RPA, using the parameters adjusted such that the RIXS data at $\delta = 0.16$ are matched.

diagram, as anticipated in Ref. [75]. Figure S3 presents a revised RPA calculation with parameters determined to fit the data at $\delta = 0.16$ ($t = 0.25$ eV, $t_z/t = 0.005$, and $\Gamma/t = 0.02$, with the remaining parameters unchanged). This parameter set recovers good agreement at $\delta = 0.16$, confirming that the RPA should not be dismissed as a practical tool for describing plasmons at any given doping, provided the band parameters are appropriately adjusted [54]. However, the same parameters cannot describe the dispersion across the other dopings, exposing the fundamental limitation of the RPA approach: the inability to transfer a single parameter set across different doping regimes. By contrast, the $t$-$J$-$V$ results presented in the main text agree well with the experimental data over the full doping and momentum range using a single set of microscopic parameters, consistent with the fact that the plasmon energy scale is governed by the correlation-renormalized electronic structure rather than by the bare band parameters alone.

## TEMPERATURE DEPENDENCE OF THE PLASMON DISPERSION

In the main text, all $t$-$J$-$V$ model calculations were performed at the temperature corresponding to the respective RIXS experiment for each doping. Here we examine the temperature dependence of the computed plasmon dispersion in more detail.

RIXS studies on LSCO found that the plasmon excitation exhibits only a marginal temperature dependence for doping levels $\delta = 0.16$, 0.2, 0.3, and 0.35 in the temperature range between 15 and 300 K [51, 75]. Within the $t$-$J$-$V$ model, previous calculations near optimal doping and within a similar temperature range have also shown weak sensitivity of the plasmon energy [63]. However, the situation is qualitatively different at low doping. Figure S4 shows the computed plasmon dispersion along the in-plane $H$ direction for several dopings between $\delta = 0.05$ and 0.40 at representative temperatures. For low doping levels, the plasmon energy exhibits an appreciable temperature dependence compared to the optimal and overdoped cases, with the plasmon energy decreasing upon increasing temperature, and the effect being most pronounced at large $H$. This enhanced sensitivity may be understood from the fact that at low doping the correlation-renormalized bandwidth, which scales as $\delta/2$, is narrow, so that thermal effects produce a stronger shift of the plasmon pole in $D_{11}(\mathbf{q},\omega)$. For $\delta \geq 0.16$, the effective bandwidth is sufficiently large so that temperature variations induce only negligible changes in the plasmon dispersion, consistent with the experimental observations in Refs. [51, 75].

---

* yamase.hiroyuki@nims.go.jp
† m.hepting@fkf.mpg.de

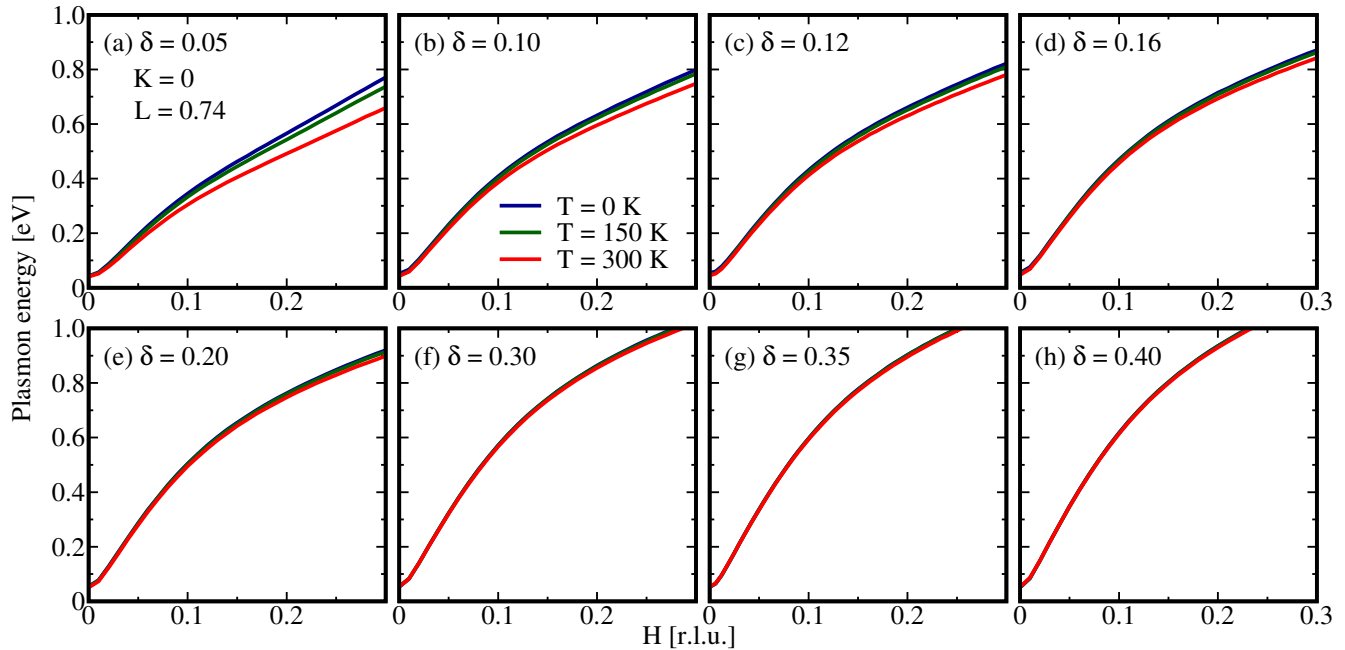

FIG. S4. Temperature dependence of the computed plasmon dispersion along the in-plane $H$ direction for several dopings between $\delta = 0.05$ and $0.40$ at fixed $K = 0$ and $L = 0.74$. Solid lines correspond to the plasmon dispersion calculated with the $t$-$J$-$V$ model. The same microscopic parameter set as in the main text is employed. Blue, green and red lines are for $T = 0$ K, $T = 150$ K, and $T = 300$ K, respectively.



\end{document}